\documentclass[useAMS,usenatbib]{mn2e}
\usepackage{graphicx}
\usepackage{lscape}

\title[A very cool binary]{The discovery of a very cool binary system}
\author[Ben Burningham et al]{Ben Burningham$^{1}$\thanks{E-mail:
    B.Burningham@herts.ac.uk}, S. K. Leggett$^{2}$, P.W. Lucas$^{1}$,
  D.J. Pinfield$^{1}$, R.L. Smart$^{3}$,  
\newauthor
A.C. Day-Jones$^{4}$, H.R.A. Jones$^1$, D.Murray$^1$
E. Nickson$^{5,1}$, M. Tamura$^{6}$,
\newauthor
 Z. Zhang$^{1}$, N. Lodieu$^{7}$, C.G. Tinney$^{8}$, M. R. Zapatero Osorio$^{9}$ \\
$^{1}$ Centre for Astrophysics Research, Science and Technology
Research Institute, University of Hertfordshire, Hatfield AL10 9AB \\
$^{2}$ Gemini Observatory, 670 N. A'ohoku Place, Hilo, HI 96720, USA \\
$^{3}$ Istituto Nazionale di Astrofisica, Osservatorio Astronomico di Torino, Strada Osservatrio 20, 10025 Pino Torinese, Italy \\
$^{4}$ Universidad de Chile,Camino el Observatorio \# 1515, Santiago, Chile, Casilla 36-D\\
$^{5}$ University of Southampton, Southampton, UK \\
$^{6}$ National Astronomical Observatory, Mitaka, Tokyo 181-8588\\
$^{7}$ Instituto de Astrof\'isica de Canarias, 38200 La Laguna, Spain\\
$^{8}$ School of Physics, University of New South Wales, 2052. Australia\\ 
$^{9}$ Centro de Astrobiolog\'\i a (CSIC-INTA), E-28850 Torrej\'on de
Ardoz, Madrid, Spain\\
}

\begin{document}
%
%
%
%


\def\aj{\rm{AJ}}                   
\def\araa{\rm{ARA\&A}}             
\def\apj{\rm{ApJ}}                 
\def\apjl{\rm{ApJ}}                
\def\apjs{\rm{ApJS}}               
\def\ao{\rm{Appl.~Opt.}}           
\def\apss{\rm{Ap\&SS}}             
\def\aap{\rm{A\&A}}                
\def\aapr{\rm{A\&A~Rev.}}          
\def\aaps{\rm{A\&AS}}              
\def\azh{\rm{AZh}}                 
\def\baas{\rm{BAAS}}               
\def\jrasc{\rm{JRASC}}             
\def\memras{\rm{MmRAS}}            
\def\mnras{\rm{MNRAS}}             
\def\pra{\rm{Phys.~Rev.~A}}        
\def\prb{\rm{Phys.~Rev.~B}}        
\def\prc{\rm{Phys.~Rev.~C}}        
\def\prd{\rm{Phys.~Rev.~D}}        
\def\pre{\rm{Phys.~Rev.~E}}        
\def\prl{\rm{Phys.~Rev.~Lett.}}    
\def\pasp{\rm{PASP}}               
\def\pasj{\rm{PASJ}}               
\def\qjras{\rm{QJRAS}}             
\def\skytel{\rm{S\&T}}             
\def\solphys{\rm{Sol.~Phys.}}      
\def\sovast{\rm{Soviet~Ast.}}      
\def\ssr{\rm{Space~Sci.~Rev.}}     
\def\zap{\rm{ZAp}}                 
\def\nat{\rm{Nature}}              
\def\iaucirc{\rm{IAU~Circ.}}       
\def\aplett{\rm{Astrophys.~Lett.}} 
\def\apspr{\rm{Astrophys.~Space~Phys.~Res.}}
\def\bain{\rm{Bull.~Astron.~Inst.~Netherlands}} 
\def\fcp{\rm{Fund.~Cosmic~Phys.}}  
\def\gca{\rm{Geochim.~Cosmochim.~Acta}}   
\def\grl{\rm{Geophys.~Res.~Lett.}} 
\def\jcp{\rm{J.~Chem.~Phys.}}      
\def\jgr{\rm{J.~Geophys.~Res.}}    
\def\jqsrt{\rm{J.~Quant.~Spec.~Radiat.~Transf.}}
\def\memsai{\rm{Mem.~Soc.~Astron.~Italiana}}
\def\nphysa{\rm{Nucl.~Phys.~A}}   
\def\physrep{\rm{Phys.~Rep.}}   
\def\physscr{\rm{Phys.~Scr}}   
\def\planss{\rm{Planet.~Space~Sci.}}   
\def\procspie{\rm{Proc.~SPIE}}   

\let\astap=\aap
\let\apjlett=\apjl
\let\apjsupp=\apjs
\let\applopt=\ao

\maketitle

\begin{abstract}
We report the discovery of a very cool d/sdL7+T7.5p common proper motion
binary system, SDSS~J1416+13AB, found by cross-matching the UKIDSS
Large Area Survey Data 
Release 5 against the Sloan Digital Sky Survey Data Release 7.
The d/sdL7
is blue in J-H and H-K and has other features suggestive of
low-metallicity and/or high gravity. 
The T7.5p displays spectral peculiarity seen before in earlier type
dwarfs discovered in UKIDSS LAS DR4, and referred to as CH$_4$-J-early
peculiarity, where the CH$_4$-J index, based on the absorption to the red
side of the $J$-band peak, suggests an earlier spectral
type than the H$_2$O-J index, based on the blue side of the $J$-band
peak, by $\sim 2$ subtypes. 
We suggest that
CH$_4$-J-early peculiarity arises from low-metallicity and/or high-gravity,
and speculate as to its use for classifying T~dwarfs.
UKIDSS and follow-up UKIRT/WFCAM photometry shows the T~dwarf to have the
bluest near-infrared colours yet seen for such an object  with
$H-K = -1.31 \pm 0.17$. 
Warm {\it Spitzer} IRAC photometry shows the T~dwarf to have extremely
red $H - [4.5]~=~4.86 \pm 0.04$, which is the reddest yet seen for a
substellar object. 
The lack of parallax measurement for the pair limits our ability to
estimate parameters for the system.
However, applying a conservative
distance estimate of 5--15~pc suggests a projected separation in range
45--135~AU. 
By comparing $H - K:H - [4.5]$ colours of the T~dwarf to spectral models we
estimate that $T_{\rm eff} = 500$~K and [M/H]$\sim -0.30$, with
$\log g \sim 5.0$. 
This suggests a mass of $\sim$30~M$_{Jupiter}$ for the T dwarf and an
age of $\sim$10~Gyr for the system.
The primary would then be a 75~M$_{Jupiter}$
object with $\log~g \sim 5.5$ and a relatively dust-free $T_{\rm
  eff} \sim 1500$K atmosphere.
Given the unusual properties of the system we caution that these
estimates are uncertain. We eagerly await parallax
measurements and high-resolution imaging which will constrain the
parameters further.

\end{abstract}


\section{Introduction}
\label{sec:intro}
The current generation of wide-field surveys \citep[e.g. UKIRT
  Infrared Deep Sky Survey, UKIDSS; Canada-France Brown Dwarf Survey,
  CFBDS; ][]{ukidss,cfbds} is significantly expanding the
sample of late type T~dwarfs \citep[e.g.][]{delorme08,lod07,pinfield08,ben10}.
Recent discoveries of extremely cool
T~dwarfs probe new low-temperature extremes, with $T_{\rm eff}$
as low as 500K \citep{ben08,delorme08,ben09,sandy09}. 
In addition to probing new $T_{\rm eff}$ regimes, we can expect
the expanded sample to populate other hitherto unexplored regions of
T~dwarf parameter space. 
Of particular interest is the growing diversity seen in metallicity
and gravity for late-T~dwarfs \citep[e.g.][]{sandy10}, and the
potential for extending the low-metallicity subdwarf sequence to very
low temperatures.

To date, the sample of ultracool subdwarfs (UCSDs) consists of just one
proposed T~subdwarf, 2MASS~J09373487+2931409
\citep{burgasser02,burgasser06}, along with a small number of
L~subdwarfs \citep[e.g. 2MASS~J1626+3925 - sdL4; SDSS~J1256--0224 -
  sdL4; 2MASS~J0616--6407 - sdL5; ULAS~J1350+0815 - sdL5; 2MASS~J0532+8246 - sdL7;][ respectively]{burgasser04a,sivarani09,cushing09,lodieu2009,burgasser03}. 
Recent parallax determinations and model comparisons
by \citet{schilbach09} suggest that of these, only the earliest type
objects (2MASS~J1626+39 and
SDSS~J1256-02) have metallicities consistent with subdwarf
classification on the scheme that \citet{gizis97} defined for
M~subdwarfs.
Based on this \citet{schilbach09} suggest that an intermediate d/sd
classification should be applied to the two coolest  objects
(2MASS~J0532+82 and 2MASS~J0937+29).  
It is important to remember, however, that the subdwarf classification
scheme is empirically based, and metallicities are associated with
specific subdwarf classes only by model comparisons. 
That the model comparisons for the latest type UCSDs suggest higher
metallicities than seen for earlier type objects should not be a sole basis
for reclassification.
The higher metallicity inferred from the colours of the coldest objects
may actually highlight problems with the models to which they are compared.

The spectral classification of subdwarfs should be
based on observed spectral features that distinguish these objects
from ``normal'' ultracool dwarfs (UCDs).
As such, in this paper we adopt the position that the sdL~objects
described above are subdwarfs, since their spectra are clearly
distinct from those of the bulk population of L~dwarfs in a manner
broadly consistent with subdwarf status.  
The more limited sample of T~dwarfs, however, precludes such
classification at this time, and we adopt the ``peculiar'' description
for possible subdwarfs of this type \citep[e.g.][]{burgasser06,ben10}. 
However, in both cases the limited sample of ``subdwarf'' objects
means that the current classification system may require significant
revision as the true diversity of the spectra of low-metallicity
UCDs becomes apparent in the era of larger, deeper surveys such
as VISTA and WISE.

We report here the discovery of a nearby d/sdL7+T7.5p common proper
motion binary. 
The rest of the paper is laid out as follows.
In Section~\ref{sec:ident} we describe the
identification, photometric follow-up, spectral classification and
proper motion determination for the two objects. In
Section~\ref{sec:binary} we demonstrate their association as a common
proper motion binary pair, and we provide initial estimates for some
of their properties in Section~\ref{sec:properties}. 
Our results and conclusions are summarised in Section~\ref{sec:summ}.

\section{Two new ultracool dwarfs}
\label{sec:ident}

Our searches of the UKIDSS Large Area
Survey \citep[LAS; see][]{ukidss} have been successful at identifying
late-type T~dwarfs
\citep[e.g.][]{lod07,warren07,pinfield08,ben08,ben09,ben10}. 
Using the same search methodology as previously described in detail in
\citet{pinfield08} and \citet{ben10}, we identified
ULAS~J141623.94+134836.30 (hereafter ULAS~J1416+13) as a candidate
late-T~dwarf in Data Release 5 of the LAS with unusually blue $H-K
=-1.35$.
The subsequent photometric and spectroscopic follow-up, which resulted in its
classification as a T7.5p dwarf, are described in the following sub-sections.

Inspection of the surrounding field in SDSS, required to establish the
red
nature of ULAS~J1416+13, revealed the presence of a nearby, very red
object at a separation of 9\arcsec.
Interrogation of SDSS DR7 revealed this object, SDSS
J141624.08+134826.7 (hereafter SDSS~J1416+13), to have an SDSS
spectrum with L~dwarf spectral morphology (see also
Table~\ref{tab:optmags} for SDSS photometry of this object).

\begin{figure}
\begin{center}
\includegraphics[height=200pt, angle=0]{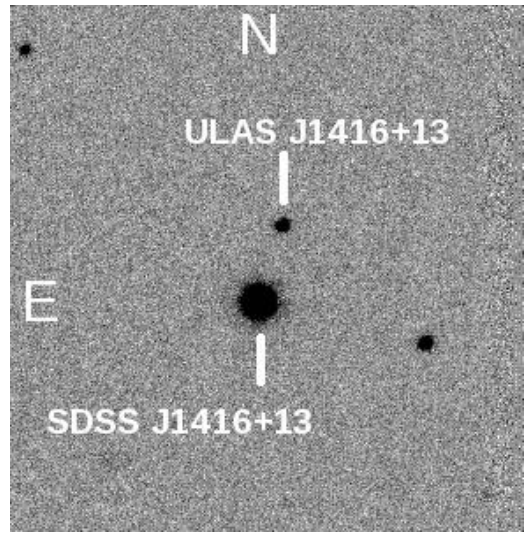}
\caption{A 1'$\times$1' $J$ band finding chart for ULAS~J1416+13
and SDSS~J1416+13 taken from the UKIDSS database.
}
\label{fig:finder}
\end{center}
\end{figure}

Since our initial identification of this L~dwarf, its discovery has
been published by \citet{schmidt10} and \citet{bowler10}, who have
classified it as a blue L5 and L6pec~$\pm 2$ dwarf respectively.
In the following sub-sections, we also describe our follow-up photometry
of this target, and describe our analysis of this source that was
carried out independently prior to the \citet{schmidt10} and
\citet{bowler10} publications.
Figure~\ref{fig:finder} shows a UKIDSS $J$ band finding chart for both
the L~and T~dwarf.

\subsection{Near-infrared photometry}
\label{subsec:photo}

Near-infrared follow-up photometry was obtained using the Wide Field
CAMera \citep[WFCAM; ][]{wfcam} on UKIRT on the night of 17$^{th}$
June 2009, and 
the data were processed using the WFCAM science pipeline by the
Cambridge Astronomical Surveys Unit (CASU) \citep{irwin04}, and
archived at the WFCAM Science Archive \citep[WSA; ][]{wsa}.
Observations consisted of a three point jitter pattern in the $Y$ and
$J$ bands, and five point jitter patterns in the $H$ and $K$ bands
repeated twice, all with 2x2 microstepping and individual exposures of
10 seconds resulting in total integration times of 120 seconds in $Y$
and $J$ and 400 seconds in $H$ and $K$. The resulting photometry for
both our targets is given in Table~\ref{tab:nirmags}.
The WFCAM filters are on the Mauna Kea Observatories (MKO) photometric
system \citep{mko}

\begin{table*}
\begin{tabular}{| c | c c c c c c c c |}
  \hline
Object & $u'$ & $g'$ & $r'$ & $i'$ & $z'$ & $g'-r'$ & $r'-i'$ & $i'-z'$    \\
\hline
 SDSS~J1416+13  & $23.55 \pm 0.57$ & $23.08 \pm 0.18$  & $20.69 \pm 0.04$  & $18.38 \pm 0.01$  & $15.92 \pm 0.01$  & $2.39 \pm 0.19$  & $2.31 \pm 0.04$ & $2.46 \pm 0.01$ \\
\hline
\end{tabular}
\caption{SDSS DR7 AB system photometry for SDSS~J1416+13. 
\label{tab:optmags}
}

\end{table*}

\begin{table*}
\begin{tabular}{| c | c c c c c c c c c c c |}
  \hline
Object & $Y$ & $J$ & $H$ & $K$ & $Y-J$ & $J-H$ & $H-K$    \\
\hline
 ULAS~J1416+13  & $18.16 \pm 0.02$ &  $17.26 \pm 0.02$  & $17.58 \pm 0.03$  & $18.93 \pm 0.24$  & $0.90 \pm 0.03$  & $-0.32 \pm 0.03$  & $-1.35 \pm 0.25$ \\
            & $18.13 \pm 0.02$ &  $17.35 \pm 0.02$  & $17.62 \pm 0.02$  & $18.93 \pm 0.17$  & $0.78 \pm 0.03$  & $-0.27 \pm 0.03$  & $-1.31 \pm 0.17$ \\
 SDSS~J1416+13  & $14.25 \pm 0.01$ &  $12.99 \pm 0.01$  & $12.47 \pm 0.01$  & $12.05 \pm 0.01$  & $1.26 \pm 0.01$  & $0.52 \pm 0.01$  & $0.42 \pm 0.01$ \\
            & $14.28 \pm 0.01$ &  $13.04 \pm 0.01$  & $12.49 \pm 0.01$  & $12.08 \pm 0.01$  & $1.24 \pm 0.01$  & $0.55 \pm 0.01$  & $0.41 \pm 0.01$ \\
           
\hline
\end{tabular}
\caption{UKIDSS LAS DR5 and WFCAM follow-up photometry of the newly
  discovered ultracool dwarfs presented here. In the case of each
  object the first row is the UKIDSS survey photometry, and the second
  is the results of the WFCAM follow-up. All near-infrared photometry
  presented here is on the Vega system and uses the MKO photometric system.
\label{tab:nirmags}
}

\end{table*}

\subsection{Warm-Spitzer IRAC photometry}
\label{subsec:irac}

The {\it Spitzer}  General Observer program 60093 allowed us to obtain
IRAC photometry of apparently very late-type T dwarfs discovered in
the UKIDSS data. This Cycle 6 warm mission program provides only
photometry at the shortest two wavelengths, [3.6] and [4.5]. Note that
[3.6] and [4.5] are nominal filter wavelengths and, as the photometry
is not colour-corrected for the dwarfs' spectral shapes, the results
cannot be translated to a flux at the nominal wavelength \citep[e.g.][]{cushing08,reach05}.

Data were obtained for SDSS~J1416+13 and ULAS~J1416+13 on
23$^{rd}$ August 2009. The telescope was pointed mid-way between the L
and T dwarf; with a separation of 9 \arcsec\ both dwarfs were near
the centre of the 5.2 arcminute field of view. 
Individual frame times were  30 seconds, repeated three
times, with a 16 position spiral dither pattern, for a total
integration time of 24 minutes in each band. The
post-basic-calibrated-data (pbcd) mosaics generated by the {\it
  Spitzer} pipeline were used to obtain aperture photometry. The
photometry was derived using a 0.6-arcsecond pixel aperture radius,
with separate (i.e. not annular) skies chosen to avoid the flaring due
to the bright primary. The aperture correction was taken from the IRAC
handbook\footnote{http://ssc.spitzer.caltech.edu/irac/dh/}. 
The error is estimated by the variation with sky aperture, which is
larger than  that implied by the uncertainty images (noise pixel maps)
provided by the {\it Spitzer} pipeline, and is much less than 1\% for
the A component in both bands, and 4\% and 0.7\% for the B component
at [3.6] and [4.5] respectively. 
The description of the primary issues with early release warm IRAC
data\footnote{http://ssc.spitzer.caltech.edu/irac/documents/iracwarmdatamemo.txt}
indicates that the only significant concern is the uncertainty in the
linearity correction for SDSS~J1416+13; the total uncertainty due to
this correction is estimated to be 5--7\% at [3.6] and 4\% at [4.5]
for bright sources. Otherwise the photometry for both sources is
uncertain by the usual 3\% due to uncertainties in the absolute
calibration and pipeline processing.  Table~\ref{tab:irac} gives the
photometry and the total uncertainties for both dwarfs.

\subsection{Near-infrared spectroscopy}
\label{subsec:spectra}

We used $JH$ and $HK$ grisms in the InfraRed Camera and Spectrograph \citep[IRCS;][]{IRCS2000} on the
Subaru telescope on Mauna Kea to obtain a R$\sim 100$ $JH$ and $HK$ spectra for
ULAS~J1416+13 on 7$^{th}$ May 2009 and 31$^{st}$ December 2009 respectively.
The observations were made up of a set of eight 300s sub-exposures for
the $JH$ spectrum and eighteen 200s sub-exposures in an ABBA jitter
pattern to facilitate effective background subtraction, with a slit
width of 1 arcsec.  
The length of the A-B jitter was 10 arcsecs.
The spectrum was extracted using standard IRAF
packages. The AB pairs were subtracted using generic IRAF tools,
and median stacked.  
The data were found to be sufficiently uniform in
the spatial axis for flat-fielding to be neglected.

We used a comparison argon arc frame to obtain the dispersion
solution, which was then applied to the pixel coordinates in the
dispersion direction on the images. 
The resulting wavelength-calibrated subtracted pairs had a low-level
of residual sky emission removed by fitting and subtracting this
emission with a set of polynomial functions fit to each pixel row
perpendicular to the dispersion direction, and considering pixel data
on either side of the target spectrum only. 
The spectra were then extracted using a linear aperture, and cosmic
rays and bad pixels removed using a sigma-clipping algorithm.

Telluric correction was achieved by dividing each extracted target
spectrum by that of the F4V star HIP 72303, which was observed just
after the target and at a similar airmass. 
Prior to division, hydrogen lines were removed from the standard star
spectrum by interpolating the stellar
continuum.
Relative flux calibration was then achieved by multiplying through by a
blackbody spectrum of the appropriate $T_{\rm eff}$.
The spectra were then normalised
using the measured near-infrared photometry to place the spectra on an
absolute flux scale, and rebinned by a factor of three to increase the
signal-to-noise, whilst avoiding under-sampling of the spectral resolution.

\subsection{Spectral types}
\label{subsec:sptypes}

As noted in the Section~\ref{sec:intro}, the discovery SDSS~J1416+13
has recently been published by \citet{schmidt10} and \citet{bowler10}.
\citet{schmidt10}  find an optical spectral type of L5 and an
infrared type of L5--6 \citep[using the ][ indices]{geballe02}. 
\citet{bowler10} similarly find an optical type of L6$\pm$0.5 and an
infrared type of L7--7.5.  
The template fits carried out in both papers show some discrepancies
beyond 9000\AA\ however, and here we use the SDSS spectrum of the
source to produce an alternative classification as follows.

\begin{figure*}
\includegraphics[height=300pt, angle=90]{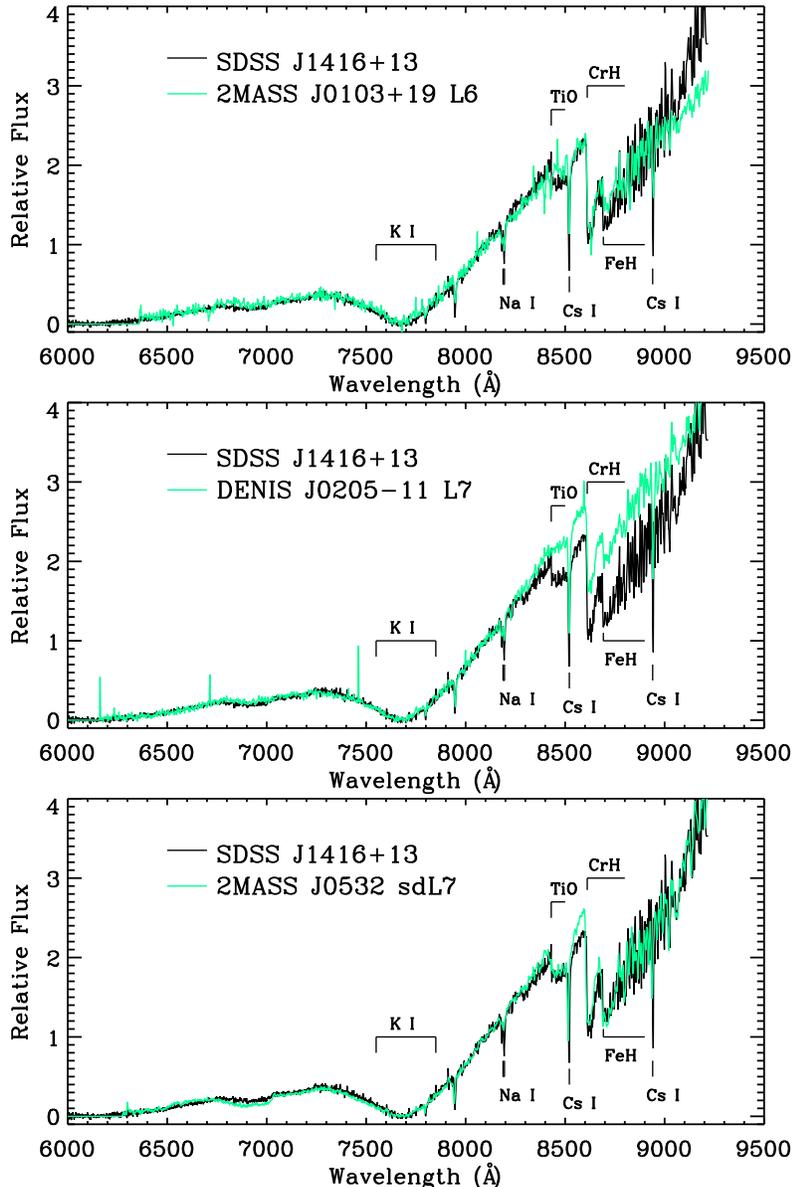}
\caption{The SDSS optical spectrum for SDSS~J1416+13 displayed
  with the template spectra for the L6 dwarf 2MASS~J0103+19
  \citep{kirkpatrick2000}, L7 dwarf DENIS~J0205--1159
  \citep{delfosse97} and sdL7 dwarf 2MASS~J0532+82
  \citep{burgasser03}. The spectra are normalised to unity at 8100\AA.}
\label{fig:sdssspec}
\end{figure*}

The top two panels of Figure~\ref{fig:sdssspec} show the SDSS DR7 spectrum
of SDSS~J1416+13 along with the optical spectra of the L6 and L7
spectral templates 2MASS~J0103+19 and DENIS~J0205--11.  
Whilst the SDSS~J1416+13 is good match over much of the range to the L6
template, they disagree significantly beyond 9000\AA.
On the other hand, the slope of the pseudo-continuum is very similar
to that of an L7 across the entire 6000--9200\AA\ range,
although the prominent TiO, FeH and CrH features are
considerably stronger in the spectrum of SDSS~J1416+13.  
This behaviour is more typical of low-metallicity objects, where it
has been speculated that that the low-metallicity
atmosphere inhibits the formation of the condensate dust clouds,
allowing the opacity due to alkali and hydride species to become more apparent
\citep[e.g.][]{burgasser03,reiners06}. 
Hence, we do not classify this object following the system for
L~dwarfs defined by \citet{kirkpatrick99}, and instead rely on
comparison to other metal-poor L~dwarfs.

The lower panel of Figure~\ref{fig:sdssspec} shows the close
similarity between the spectrum of SDSS~J1416+13 and that of the
metal-poor L~dwarf 2MASS~J0532+8246.
\citet{burgasser03} demonstrated that this object not only displays
features characteristic of a low-metallicity atmosphere, but also has
kinematics consistent with halo membership, and classify it as sdL7.
Whilst the general agreement between the spectrum
of SDSS~J1416+13 and the sdL7 spectrum is good across the entire range
considered, there are specific areas of disagreement that should be
noted. 
In particular the Cs{\sc I} and Na{\sc I} absorption features are
somewhat deeper than in the sdL7 template, and more suggestive
of dwarf classification than that of a subdwarf.
This suggests that SDSS~J1416+13 may be less metal poor than
2MASS~J0532+82. 
Given the apparent intermediate nature of SDSS~J1416+13 between the L7
and sdL7 spectra, we classify it as d/sdL7 (optical).
We note that \citet{bowler10} suggest that SDSS~J1416+13 is unlikely
to have significantly reduced metallicity based on the optical TiO and
CaH features. \citet{burgasser08b} and \citet{stephens09} discuss
various mechanisms which may lead to unusually blue L dwarfs including
low metallicity, high gravity and thin condensate cloud decks. We
explore the physical properties of the L dwarf further in
Section~\ref{sec:properties}.

\begin{figure*}
\includegraphics[height=500pt, angle=90]{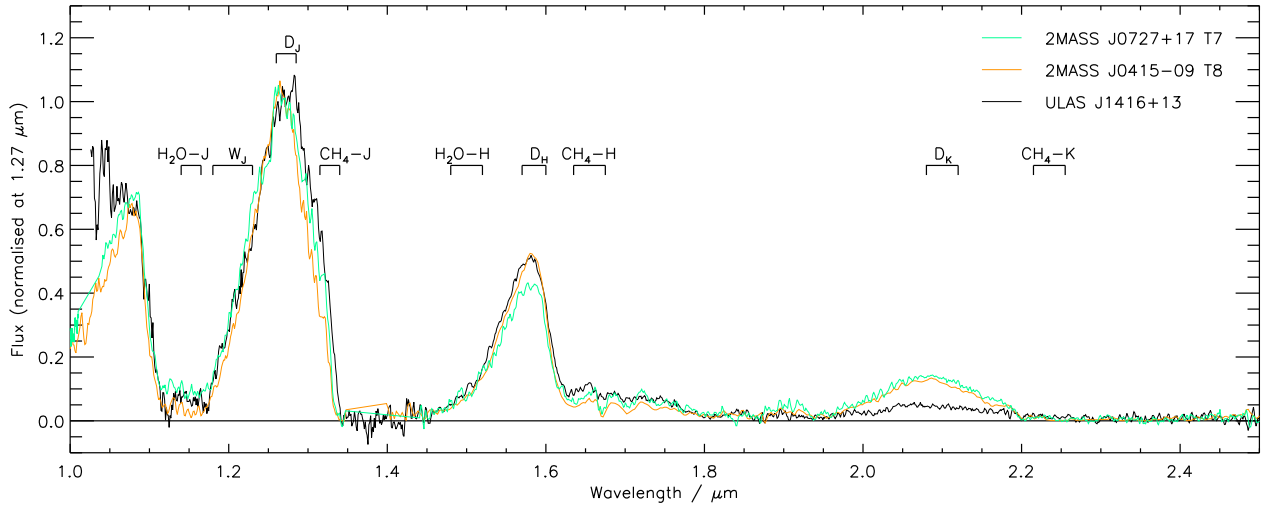}
\caption{The IRCS $JHK$ spectrum for ULAS~J1416+13 plotted with T7 and
  T8 spectral standards 2MASS~J072718.24+171001.2 and 2MASS
  J04151954--093506.6. Overlaid are the spectral ranges for the
  numerators and denominators (D) of the spectral typing indices of
  \citet{burgasser06} and \citet{ben08}.}
\label{fig:JHKspec}
\end{figure*}

The IRCS spectrum of the T~dwarf, ULAS~J1416+13, is shown in
Figure~\ref{fig:JHKspec}, along with spectra of the T7 and T8 spectral
standards \citep{burgasser06}.
With the exception of the poor match to both templates on the red side
of the J-band peak and the heavily suppressed $K$ band peak, the
spectrum appears intermediate between the two. 
This is reflected in the spectral typing ratios (see
Table~\ref{tab:indices}), and we classify this object as T7.5p.
The early type suggested by the CH$_{4}$-K index clearly reflects the
small amount of flux in the $K$~band peak.
The type of peculiarity seen here in the red side of the $J$~band peak,
and reflected in the spectral typing ratios, has been described for at
least three other T~dwarfs in \citet{ben10}, and has been
suggested as a possible tracer of low-metallicity and/or
high-gravity. The significance of this feature is discussed in more
detail in Section~\ref{sec:properties}. 

\begin{table}
\begin{tabular}{| c c c |}
  \hline
Name      &        [3.6] &   [4.5] \\
\hline
SDSS J1416+13  &   $10.99 \pm 0.07$ &  $10.98 \pm 0.05$ \\
ULAS J1416+13  &   $14.69 \pm 0.05$ &  $12.76 \pm 0.03$ \\

\hline
\end{tabular}
\caption{{\it Spitzer} IRAC photometry for the d/sdL7 and T7.5p dwarfs
  presented here.
\label{tab:irac}}

\end{table}

\begin{table}\renewcommand{\arraystretch}{3}\addtolength{\tabcolsep}{-1pt}
\begin{tabular}{c c c c }
  \hline
 {\bf Index} & {\bf Ratio} & {\bf Value} & {\bf Type} \\
\hline
H$_2$O-J & $\frac{\int^{1.165}_{1.14} f(\lambda)d\lambda}{\int^{1.285}_{1.26}f(\lambda)d\lambda }$ & $0.07 \pm 0.01$  & T7/8 \\[+1mm]
CH$_4$-J & $\frac{\int^{1.34}_{1.315} f(\lambda)d\lambda}{\int^{1.285}_{1.26}f(\lambda)d\lambda }$ &  $0.34 \pm 0.01$ & T6 \\
$W_J$ & $\frac{\int^{1.23}_{1.18} f(\lambda)d\lambda}{2\int^{1.285}_{1.26}f(\lambda)d\lambda }$   &  $0.34 \pm 0.01$  & T7/8 \\
H$_2$O-$H$ & $\frac{\int^{1.52}_{1.48} f(\lambda)d\lambda}{\int^{1.60}_{1.56}f(\lambda)d\lambda }$ & $0.20 \pm 0.01$  & T7/8 \\
CH$_4$-$H$ & $\frac{\int^{1.675}_{1.635} f(\lambda)d\lambda}{\int^{1.60}_{1.56}f(\lambda)d\lambda }$ & $0.20 \pm 0.01$ & T7 \\
NH$_3$-$H$ &  $\frac{\int^{1.56}_{1.53} f(\lambda)d\lambda}{\int^{1.60}_{1.57}f(\lambda)d\lambda }$ & $0.61 \pm 0.01$ & ... \\
CH$_4$-K &  $\frac{\int^{2.255}_{2.215} f(\lambda)d\lambda}{\int^{2.12}_{2.08}f(\lambda)d\lambda }$ & $0.29 \pm 0.02$ & T4 \\
\hline
\end{tabular} 
\caption{The spectral flux ratios for ULAS~J1416+13. Those used for
  spectral typing are indicated on Figure~\ref{fig:JHKspec}.The NH$_3$
index is not used for assigning a type \citep[see ][ and Burningham et
al 2010 for a discussion of this]{ben08}, but is included for
completeness and to permit future comparison with other late T~dwarfs. }
\label{tab:indices}
\end{table}

\subsection{Proper motions}
\label{subsec:propermotion}

The photometric follow-up observations that were carried out provided
a second epoch of imaging data, showing the 
position of the two sources of interest 1.1 years after the LAS image
was measured. 
We used the {\scriptsize IRAF} task {\scriptsize GEOMAP} to derive
spatial transformations from the WFCAM follow-up $J$-band image into
the original UKIDSS LAS $J$-band image based on the positions of 18
reference stars.  
The transform allowed for linear shifts and rotation, although the
rotation that was required was negligible.
We then transformed the WFCAM follow-up pixel coordinates of the targets
into the LAS images using {\scriptsize GEOXYTRAN}, and calculated
their change in position (relative to the reference stars) between the
two epochs.  

The uncertainties associated with our proper motion measurement
primarily come from the spatial transformations, and the accuracy with
which we have been able to measure the position of the targets (by
centroiding) in the image data. 
Centroiding uncertainties for the targets should be small, since
the seeing and signal-to-noise of the sources was
good in both epochs, so this latter source of uncertainty will be
neglected. 
For the LAS image the seeing was
$\sim$0.9\arcsec\ in the $J$-band, whilst for the WFCAM image it was
$\sim 1.1$\arcsec.
The root-mean-square (rms) scatter in
the difference between the transformed positions of the reference
stars and their actual measured positions was $\pm$0.24 pixels in
declination and $\pm$0.18 pixels in right ascension, corresponding to 0.048 and 0.036\arcsec\  in the $J$-band LAS image.
 We thus estimate proper motion uncertainties of $\pm$45 mas/yr and
 $\pm$33 mas/yr in declination and right ascension respectively. 
The final, relative, proper motion measurements are $\mu_{\alpha
  cos\delta}=248 \pm 33$mas/yr, $\mu_{\delta}= 100 \pm 45$mas/yr for
SDSS~J1416+13 and $\mu_{\alpha cos\delta}=221 \pm 33$mas/yr,
$\mu_{\delta}= 115 \pm 45$mas/yr for ULAS~J1416+13.

It should be noted that the relative proper motions calculated here disagree
with the absolute values found for the primary by \citet{schmidt10} and
\citet{bowler10} at the 4$\sigma$ level.
This discrepancy likely arises as a result of two factors. 
Firstly, in the first epoch images both targets lie within
30\arcsec\ of the detector edge.
As a result, the distribution of reference stars is not even about the
targets. 
Since there is likely to considerable geometric distortion across the
field of view, this poor distribution of reference stars will likely result
in an unreliable absolute fit to the coordinates.
Secondly, they do not take into account the parallax of the
targets. 
The first and second epoch data were taken on 12$^{th}$ May
2008 and 17$^{th}$ June 2009 respectively, which would suggest the
influence of parallax should be small. 
However, given that the distance for both objects
may be as low as 5~pc (see Section~\ref{sec:properties}), we do not
rule this out as a significant effect.
These concerns should not effect the reliability of these
proper motions as relative values, but we caution that they include
systematic effects that prevent their use in any absolute manner.

\section{A wide low-mass binary}
\label{sec:binary}

The close agreement of the proper-motions for these two objects, and
their 9 \arcsec\ proximity on the sky suggests that they represent a
common proper motion binary pair. 
To estimate the probability that the proper motions are aligned by
chance, rather than because of a bona-fide association, we have
considered the proper motions of objects in the SuperCosmos Sky Survey
\citep{supercos1} in the direction of our targets.
Since we do not have a parallax for either object, we instead estimate
a liberal range of distances based on their spectral types and
apparent magnitudes for the purposes of placing broad limits on their
shared volume. 
In Section~\ref{sec:properties} we refine this distance estimate based
on subsequent analysis of these objects. 
If we apply the $M_J$ vs. spectral type relations of \citet{liu06} we
find that an L7 and a T7.5 dwarf with the apparent magnitudes of our
objects can be expected to lie at distances ranging from 5~pc to 25~pc. 
Of the $\sim 50$ SuperCosmos objects with apparent distances (based on
colour-magnitude relation for field stars) similar to those of our
targets, none shared a common motion to within $2 \sigma$. 
We thus conclude that the likelihood of a common proper motion
occurring by chance in this direction is less than $1/50$.

Since the statistics for the
properties of the ultracool subdwarf population are not currently
known, we will use the space density of ``normal'' L~dwarfs to
estimate a conservative
probability that this pair are unrelated, and are found in close
proximity by chance.
Using our liberal distance range of 5--25~pc, and given the separation
of 9\arcsec, we can thus estimate that two objects likely share a
volume of $\leq$0.01~pc$^3$. 
The space density for field L~dwarfs was determined by \citet{cruz07} to be
0.0038~pc$^{-3}$. 
The probability of finding an L~dwarf within the same 0.01~pc$^3$ as
our T7.5 dwarf is thus $3.8 \times 10^{-5}$. 
It is reasonable to surmise that the probability of finding two
ultracool subdwarfs within this volume would be considerably smaller.

These combined arguments suggest the probability of a chance
alignment in space and motion for these two objects is less than
$10^{-6}$.
If we apply these arguments to the total UKIDSS LAS T~dwarf sample up
to DR4 \citep{ben10} we find that we would need a sample
of approximately 1000 times larger before we would expect to identify
one chance alignment such as this. 
It is worth stressing that our estimate for this probability is
somewhat conservative. 
Given the apparently unusual nature of the objects
discussed here, it is likely that true probability for chance alignment is
considerably lower.
 We thus conclude that SDSS~J1416+13 and ULAS~J1416+13 represent a
binary pair, which we shall henceforth refer to as SDSS~J1416+13AB.

\section{The properties of SDSS~J1416+13AB}
\label{sec:properties}

\begin{figure}
\includegraphics[height=300pt, angle=0]{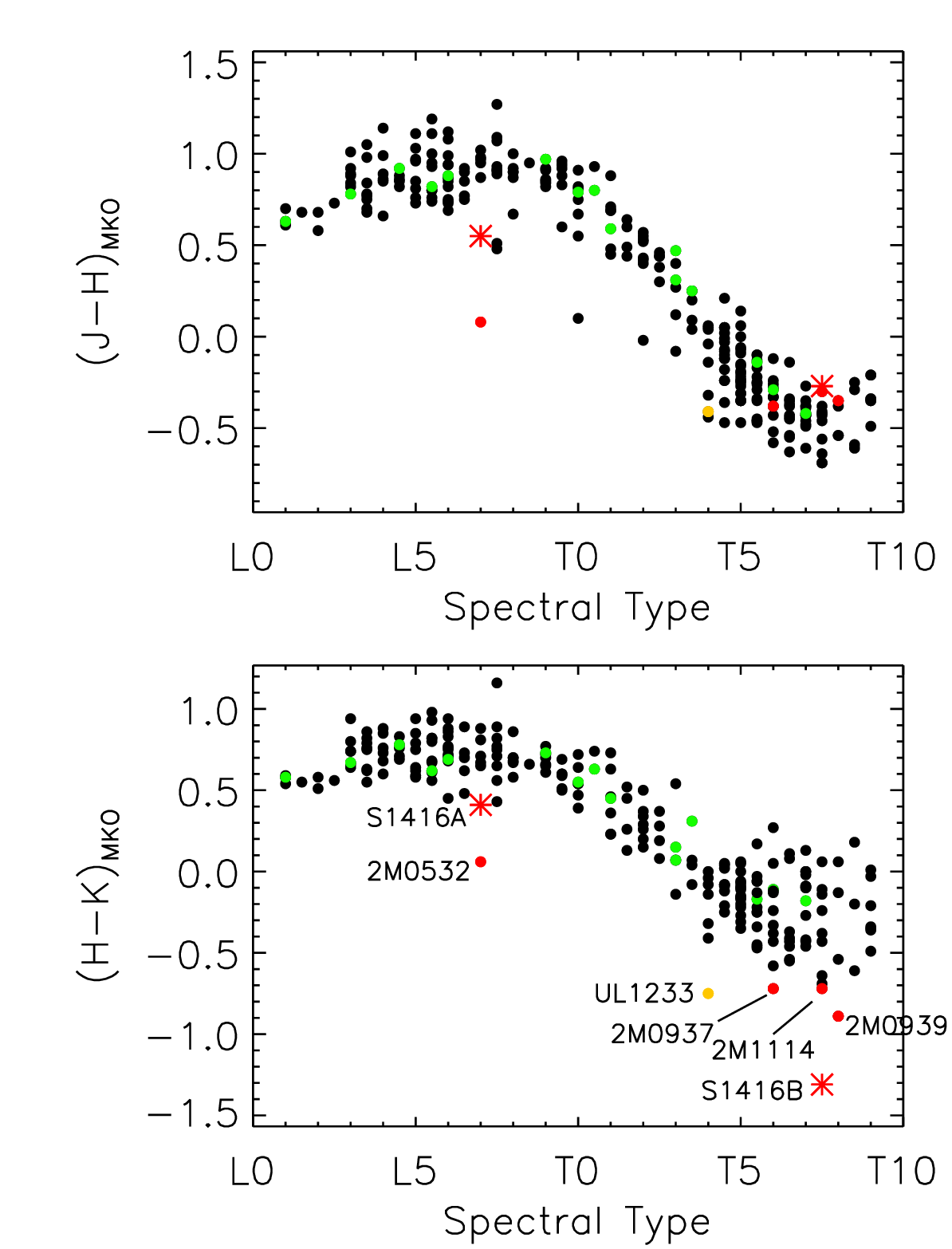}
\caption{$J-H$ and $H-K$ colour as a function of spectral type for L~and
  T~dwarfs. Data for L~and T~ dwarfs on the MKO system are taken from
  \citet{knapp04} with T~spectral types updated to the
  \citet{burgasser06} system. Additional data for late-T~dwarfs taken
  from \citet{ben10}.
Known binary systems are shown as green dots, whilst known metal poor
objects discussed in the text are shown as red dots, and labelled in
the lower plot. The only other known T~dwarf with $K$ band photometry
that displays CH$_4$-J-early peculiarity is shown as an orange dot, whilst
SDSS~J1416+13AB are shown as red asterisks. 
With the exception of SDSS~J1416+13A and 2MASS~J0532+82, all spectral
types are near-infrared types.
2MASS photometry for 2MASS0532+82 has been converted to the MKO system
using the \citet{stephens04} relationships, which give consistent
results with synthetic colours calculated from the object's
near-infrared spectrum.
}
\label{fig:colplot}
\end{figure}

The optical spectral classification of SDSS~J1416+13A as a
dwarf/subdwarf implies that we could reasonably classify the secondary
as a dwarf/subdwarf also, given that most binary systems are expected to be
coevally formed in the same cloud core.
Figure~\ref{fig:colplot} shows near-infrared colours as a function of
spectral type for L~and T~dwarfs, with SDSS~J1416+13AB indicated with
red asterisks. 
Blue $H-K$ near-infrared colours for mid-to late T~dwarfs have
typically been interpreted as indicative of low-metallicity and/or high-gravity
\citep[e.g.][]{burgasser02,knapp04,liu07}, caused by $K$~band
suppression by pressure sensitive collisionally induced absorption by
hydrogen \citep[CIA H$_2$;][]{saumon94}. 
Blue $J-H$ colours in metal poor L~dwarfs have also been interpreted
in terms of $H$ band suppression by CIA H$_2$ \citep[e.g.][]{burgasser03}.
The blue $J-H$ colour of SDSS~J1416+13A, and the blue $H-K$ colour of
SDSS~J1416+13B, therefore, support the interpretation that both
objects have low-metallicity and/or high-gravity, and we interpret the
peculiar spectral shape of SDSS~J1416+13B in this context.

The spectral morphology in the $J$ band peak of SDSS~J1416+13B is
reminiscent of a number of T~dwarfs recently discovered that have been
classified as peculiar \citep{ben10}.  
These also show a $J$ band peak that appears earlier in type on the
red side (as indicated by the CH$_4$-J index) compared to the blue
side (as indicated by the H$_2$0-J and $W_J$ indices).
This morphology was referred to by \citet{ben10} as
CH$_4$-J-early peculiarity, and we continue this convention here. 
Only one of the objects already found with CH$_4$-J-early peculiarity,
ULAS~J1233+1219, currently has $K$~band photometry.
It also appears very
blue, with $H-K = -0.75$ (indicated by an orange filled circle in
Figure~\ref{fig:colplot}), and is as notable an outlier in $H-K$ for its
type as SDSS~J1416+13AB.
It thus seems plausible that CH$_4$-J-early peculiarity is indicative of
low-metallicity and/or high gravity.

\begin{figure}
\includegraphics[height=250pt, angle=90]{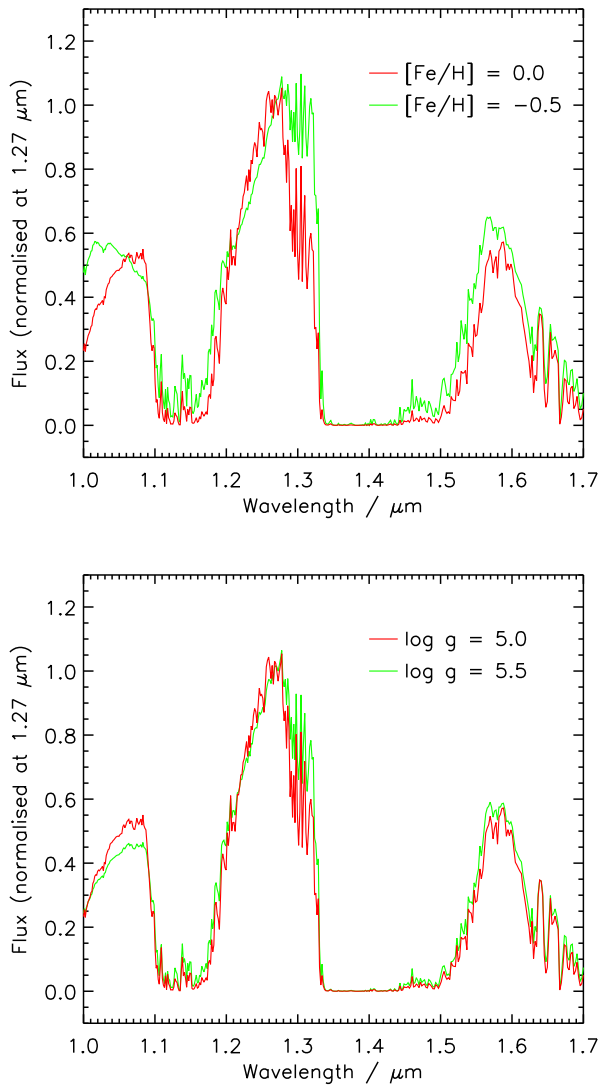}
\caption{\citet{bsh2006} models for $\log~g = 5.0$ 700K T~dwarfs with solar and
  [Fe/H]0=-0.5 metallicity, and for solar metallicity combined with
  $\log g =  5.0$ and $\log g = 5.5$.
}
\label{fig:metmod}
\end{figure}

There is some theoretical basis for preferring a low-metallicity
interpretation of CH$_4$-J-early peculiarity.
Figure~\ref{fig:metmod} shows comparisons of \citet{bsh2006} model spectra for 
$\log~g = 5.0$,  $T_{\rm eff} = 700$K T~dwarfs with solar and
[Fe/H]=-0.5 metallicity, and also for solar metallicity with $\log g =
5.0$ and $\log g = 5.5$.
Enhancement of the red side of the $J$~band peak is apparent in both
the low-metallicity and high-gravity cases, but is most pronounced in
the former.
We speculate that CH$_4$-J-early peculiarity may represent a useful
tracer of low-metallicity atmospheres, although its presence in a
system with fiducial metallicity and age constraints will need to
be observed before a robust interpretation will be possible.

It is interesting to note that the spectral shape of SDSS~J1416+13B
also deviates from that of the spectral templates blueward of
1.1$\mu$m, in a
manner similar to that seen in Figure~\ref{fig:metmod} for the
low-metallicity case. The same behaviour is not predicted for the
high-gravity case. A spectrum with better coverage in the
$Y$ band may provide a useful means of breaking the
gravity-metallicity degeneracy.

The need for a more complex spectral classification scheme to take
account of spectral variations that result from changes in
metallicity and gravity in addition to $T_{\rm eff}$  has been
highlighted by \citet{kirkpatrick05}. 
As more objects that exhibit CH$_4$-J-early spectral peculiarity are
identified, its behaviour may provide a convenient method for more
detailed classification of T~dwarf spectra.

The lack of a known parallax for this binary pair precludes a detailed
assessment of their properties, since spectroscopic distances are not
well constrained for ultracool T~dwarfs.
In the case of the one previously identified sdL7, 2MASS~J0532+82, the
determined absolute magnitude ($M_J$ = 13.00) is 1-2 mags
brighter than might otherwise be expected for a field dwarf of type L7
\citep{burgasser08}. 
If we assume that SDSS~J1416+13A has $M_J = 13.0$ as was the case for
for 2MASS~J0532+82 we arrive at a  distance estimate of 10~pc.
However, SDSS~J1416+13A is considerably less blue in $J-H$ than
2MASS~J0532+82 ($J-H = 0.55$ vs $J-H  = 0.08$ respectively, see
Figure~\ref{fig:colplot}) and, as previously discussed, may have
rather different properties.
 
Assuming spectral types L7 and T7.5, however, suggests distances of 5~pc and
20~pc for the objects respectively by applying the $M_J$ vs spectral
type relations of \citet{liu06}.
In the case of the metal poor T~dwarf 2MASS~J0937+29 this
method would overestimate the distance by $\sim 30$\%. 
Using this as a correction suggests a distance for
SDSS~J1416+13AB of 14~pc. 
This would represent a significant discrepancy in the distances of the
primary and secondary members of SDSS~J1416+13AB, implying that the
primary could be an unresolved binary.
However, the $K$~band suppression in SDSS~J1416+13B is greater than in the case
of 2MASS~J0937+29 and, as discussed below, it appears to be
considerably cooler.  
It thus seems likely that SDSS~J1416+13B is fainter still, and a
distance as close as 10~pc seems plausible.
A distance of $\sim 10$pc is also in broad agreement with that
estimated by \citet{schmidt10} and \citet{bowler10} for the primary.
We thus conservatively estimate the distance to SDSS~J1416+13AB to lie
in the 5-15~pc range.

The implied projected separation of the binary pair at this range of
distances is 45 -- 135~AU.
It is thus possible that this pair also represents a rare very
low-mass wide binary system \citep[e.g. Figure~9 in][]{lafreniere08}.

\begin{figure}
\includegraphics[height=300pt, angle=0]{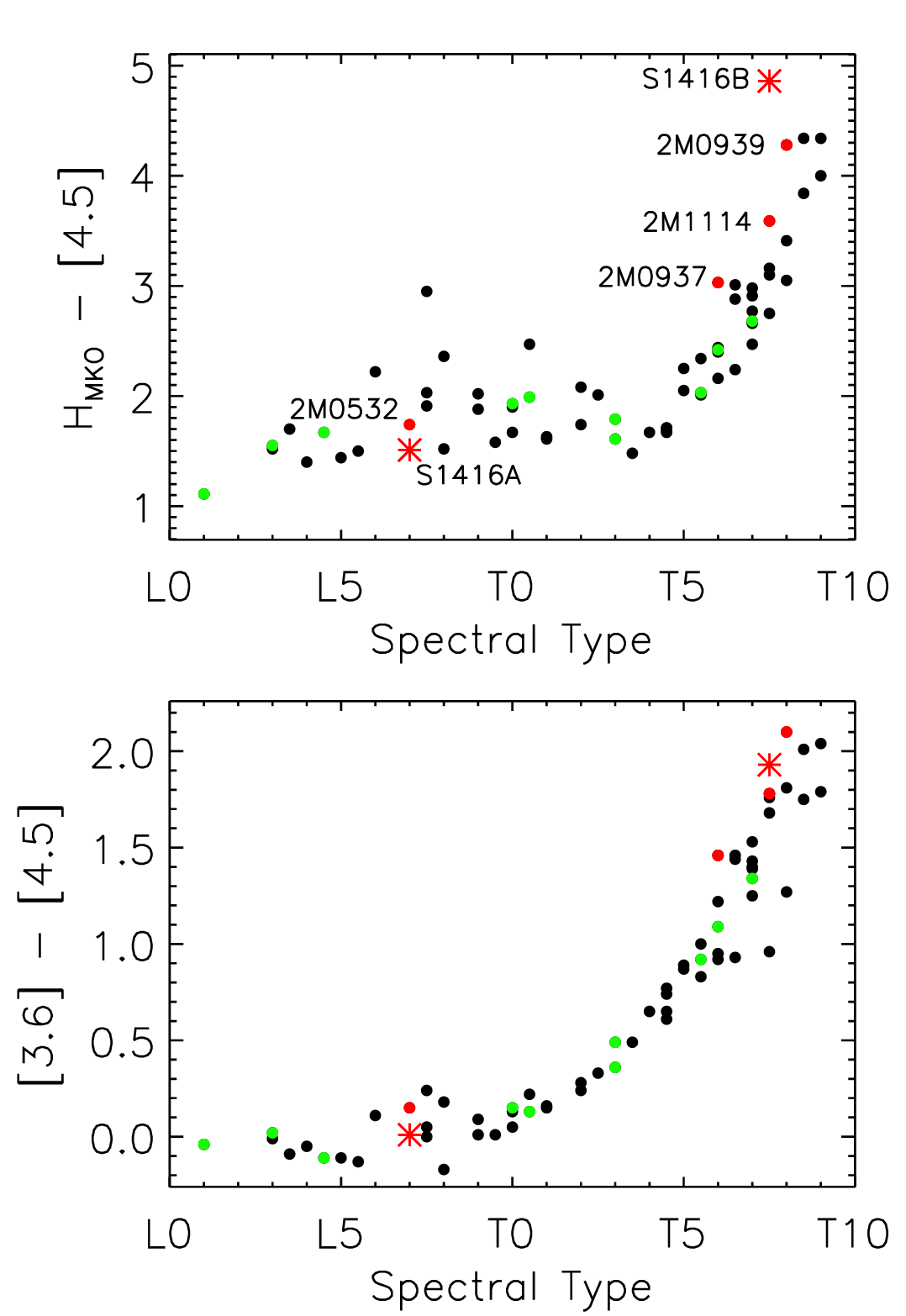}
\caption{Spitzer IRAC colours as a function of spectral type for L~and
  T~dwarfs. Data for L~and T~ dwarfs on the MKO system taken from
  \citet{knapp04} with T~spectral types updated to the
  \citet{burgasser06} system. Additional data for late-T~dwarfs taken
  from \citet{sandy10}.
Known binary systems are shown as green dots, whilst known metal poor
objects discussed in the text are shown as red dots, and labelled in
the upper plot. SDSS~J1416+13AB are shown as red asterisks. 
}
\label{fig:mircol}
\end{figure}

The longer baseline provided by our {\it Spitzer} IRAC photometry
offers the opportunity to estimate parameters of the system
through comparison to predictions of  model spectra.
The IRAC colours of SDSS~J1416+13A are normal for a late-type L dwarf,
although the colours of these objects show significant scatter (see
Figure~\ref{fig:mircol}). 
All of the low-metallicity late-T dwarfs plotted in
Figure~\ref{fig:mircol} display $H - [4.5]$ that is at least 0.5
magnitudes redder than would otherwise be expected for a ``normal''
T~dwarf of their subtype.
However, the $H_{MKO}$ - [4.5] colour of SDSS~J1416+13B is the reddest
yet measured.  
In addition to apparently indicating low-metallicity atmospheres, this
colour is a good indicator of $T_{\rm eff}$
\citep[e.g.][]{warren07,stephens09,sandy10}  and so SDSS~J1416+13B
appears to be very cool.

\begin{figure*}
\includegraphics[height=400pt, angle=270]{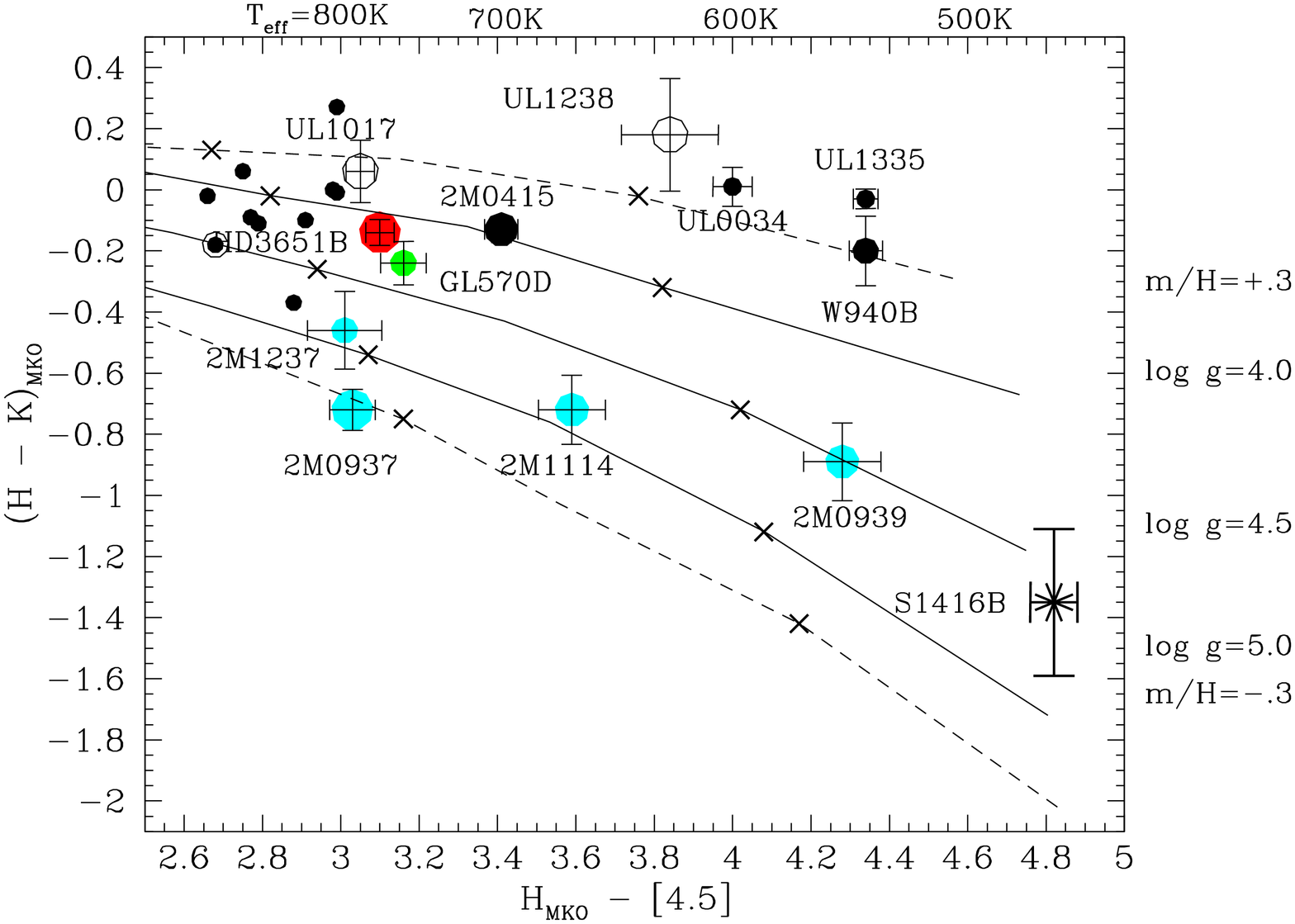}
\caption{A plot of $H-K$ vs $H_{MKO} - [4.5]$ for a selection of
  late-T dwarfs and model predictions of \citet{marley02} and
  \citet{sm08}, after Figure~11 from \citet{sandy10}. The location of SDSS~J1416+13B is indicated with an asterisk.  
The small dots represent the comparison sample of late-T dwarfs used
in \citet{sandy10}, with binaries shown as ringed symbols. Larger
symbols are T~dwarfs examined in detail in that work, where symbol
size indicates gravity and colour indicates metallicity.  Largest to
smallest filled circles correspond to $\log g  \approx 5.4$
(5.2--5.5), $\log g \approx 5.2$ (5.0 –-5.4), $\log g \approx 5.0$
(4.8 -- 5.1), and $\log g \approx 4.3$ (4.0 -- 4.5).  Open circles are
unconstrained gravity. Red symbols indicate metal-rich, green solar,
and cyan metal-poor dwarfs; black are dwarfs with unconstrained
metallicity. Model sequences with a range of gravity and metallicity
are shown, with $\log g = 4.0$, 4.5 and 5.0 and [m/H]=0 shown
as solid lines, whilst $\log g = 4.5$ with [m/H]= --0.3 and +0.3 are
shown as dashed lines. The $T_{\rm eff}$ values for the $\log g = 4.5$
[m/H]= 0 model are indicated on the top axis, and crosses along the
sequences indicate the 800K and 600K points for each model set.}
\label{fig:h449}
\end{figure*}

Figure~\ref{fig:h449} reproduces Figure~11 of \citet{sandy10} with the
location of SDSS~J1416+13B indicated.  
It can be seen that this T~dwarf
forms a sequence with the other known metal-poor ([m/H]$\sim$ -0.3)
high-gravity ($\log g \sim  5.0 - 5.3$) dwarfs: 2MASS J0937347+293142, 2MASS
J12373919+6526148, 2MASS J11145133--2618235, 2MASS J09393548--2448279.
The dwarfs have $T_{\rm eff} \sim 950, 825, 750$~and 600~K
respectively \citep{geballe09,lb07,sandy07,sandy09,burgasser08a}.
Extrapolating these values using
Figure~\ref{fig:h449} and the models \citep{marley02,sm08} shown in
the figure, implies that SDSS~J1416+13B has [m/H]$\sim$--0.3, ${\rm
  log}~g \sim$~5.0 to 5.3 and $T_{\rm eff} \sim 500$~K. This indicates
a mass of 30--40 $M_{Jupiter}$ for SDSS~J1416+13B and an age around
10~Gyr or older for the system using the evolutionary models of
\citet{sm08}.

The near-infrared indices of \citet{geballe02} for  SDSS~1416+13A
 suggest a near-infrared spectral type of L7-7.5 \citep{bowler10}.
The near-infrared spectral type-$T_{\rm eff}$
relations of \citet{stephens09} suggest that blue L7 dwarfs have
$T_{\rm eff} \sim$1500~K.   
If the system is aged at $\sim 10$Gyr as implied by the secondary then
the evolutionary models of \citet{sm08} suggest that the primary is a
$\sim$75M$_{Jupiter}$~dwarf with $\log~g \sim 5.5$.
Hence the L~dwarf is at the stellar/substellar boundary as also
suggested by \citet{bowler10}.

\citet{stephens09} have shown that the atmospheres of blue L dwarfs
are less dusty than the bulk population, deriving a high value of
$f_{sed} \sim 3$ using the \citet{marley02} models.
These authors also find that there is an indication that dust
clearing may occur at higher temperatures for higher gravity systems. 
The blue colours and almost dust-free atmosphere of this relatively
warm $\sim$1500~K~L~dwarf is therefore consistent with a high gravity
and relatively old age for the system.

Both \citet{schmidt10} and \citet{bowler10} have estimated $(U,V,W)_{\rm
  LSR}$ for SDSS~J1416+13A, finding $(-7.9 \pm 2.1, 10.2 \pm 1.2, -31.4
\pm 4.7)$~kms$^{-1}$ and $(-6 \pm 4, 10.2 \pm 1.4, -27 \pm 9)$~kms$^{-1}$
  respectively\footnote{$U$ positive towards the Galactic centre.}, 
 and interpret its kinematics as indicative of thin disk membership.
This is consistent with the age of ~10 Gyr and slight
metal-paucity that we find here \citep{robin03,haywood97}.

It is intriguing that SDSS~J1416+13B appears to be $\sim$250~K and
$\sim$100~K cooler than 2MASS~J1114--26 2MASS~J0939--24 respectively,
which are of similar spectral type (T7.5 and T8). 
It is plausible that the near-infrared spectral type
vs. $T_{\rm eff}$ relation for very late T~dwarfs shows significant
dependence on metallicity and gravity, with lower-metallicity dwarfs
of a given subtype having lower $T_{\rm eff}$ than similar type
objects of higher metallicity.
The cool nature of 2MASS~J0939--24 compared to other T8 dwarfs spectral
type would tend to support this assertion, although interpretation of
this object is complicated by its probable binarity
\citep{burgasser08a}. 
If SDSS~J1416+13B had significantly lower metallicity and/or higher gravity than
2MASS~J1114--26 2MASS~J0939--24, such an effect might account for their
similar types but diverse $T_{\rm eff}$s. 
However, the same model predictions seen in Figure~\ref{fig:h449} that
suggest such different $T_{\rm eff}$ also suggest fairly similar
metallicities and gravities for the three objects.

Finally, it should be noted that we cannot currently rule out the
presence of a cooler unresolved companion to SDSS~J1416+13B, which
might explain its extremely red $H-[4.5]$ colour coupled with is T7.5p
near-infrared morphology.
Unfortunately, the lack of parallax and high-resolution imaging for
this target prevent us from adequately exploring this issue here.

\section{Summary}
\label{sec:summ}

We have identified what appears to be the coolest binary
system yet found. 
The association of the T7.5p
component with the d/sdL7 primary allows us to now extend the
high-gravity and low-metallicity 
sequence to the lowest observed temperatures, and we suggest that CH$_4$-J-early
peculiarity \citep{ben10} may in future prove to be a useful
discriminator for this type of object.
The likely close proximity of the system to the Sun should facilitate
the determination of the trigonometric parallax in the near-future,
which will allow a more robust determination of the properties for
this exciting system.

\section*{Acknowledgements}

SKL is supported by the Gemini Observatory, which is operated by AURA,
on behalf of the international Gemini partnership of Argentina,
Australia, Brazil, Canada, Chile, the United Kingdom, and the United
States of America. 
EN received support from a Royal Astronomical Society small grant.
NL was funded by the Ram\'on y Cajal fellowship number 08-303-01-02\@.
CGT is supported by ARC grant DP0774000.
This research has made use of the SIMBAD database,
operated at CDS, Strasbourg, France, and has benefited from the SpeX
Prism Spectral Libraries, maintained by Adam Burgasser at
http://www.browndwarfs.org/spexprism.

\bibliographystyle{mn2e}
\bibliography{refs}

\begin{thebibliography}{55}
\expandafter\ifx\csname natexlab\endcsname\relax\def\natexlab#1{#1}\fi

\bibitem[{{Bowler} {et~al.}(2009){Bowler}, {Liu}, \& {Dupuy}}]{bowler10}
{Bowler} B.~P., {Liu} M.~C., {Dupuy} T.~J., 2009, ApJ in press, ArXiv 0912.3796

\bibitem[{{Burgasser} {et~al.}(2006){Burgasser}, {Geballe}, {Leggett},
  {Kirkpatrick}, \& {Golimowski}}]{burgasser06}
{Burgasser} A.~J., {Geballe} T.~R., {Leggett} S.~K., {Kirkpatrick} J.~D.,
  {Golimowski} D.~A., 2006, \apj, 637, 1067

\bibitem[{{Burgasser} {et~al.}(2002){Burgasser}, {Kirkpatrick}, {Brown},
  {Reid}, {Burrows}, {Liebert}, {Matthews}, {Gizis}, {Dahn}, {Monet}, {Cutri},
  \& {Skrutskie}}]{burgasser02}
{Burgasser} A.~J., {Kirkpatrick} J.~D., {Brown} M.~E., {Reid} I.~N., {Burrows}
  A., {Liebert} J., {Matthews} K., {Gizis} J.~E., {Dahn} C.~C., {Monet} D.~G.,
  {Cutri} R.~M., {Skrutskie} M.~F., 2002, \apj, 564, 421

\bibitem[{{Burgasser} {et~al.}(2003){Burgasser}, {Kirkpatrick}, {Burrows},
  {Liebert}, {Reid}, {Gizis}, {McGovern}, {Prato}, \& {McLean}}]{burgasser03}
{Burgasser} A.~J., {Kirkpatrick} J.~D., {Burrows} A., {Liebert} J., {Reid}
  I.~N., {Gizis} J.~E., {McGovern} M.~R., {Prato} L., {McLean} I.~S., 2003,
  \apj, 592, 1186

\bibitem[{{Burgasser} {et~al.}(2008{\natexlab{a}}){Burgasser}, {Liu},
  {Ireland}, {Cruz}, \& {Dupuy}}]{burgasser08}
{Burgasser} A.~J., {Liu} M.~C., {Ireland} M.~J., {Cruz} K.~L., {Dupuy} T.~J.,
  2008{\natexlab{a}}, \apj, 681, 579

\bibitem[{{Burgasser} {et~al.}(2008{\natexlab{b}}){Burgasser}, {Looper},
  {Kirkpatrick}, {Cruz}, \& {Swift}}]{burgasser08b}
{Burgasser} A.~J., {Looper} D.~L., {Kirkpatrick} J.~D., {Cruz} K.~L., {Swift}
  B.~J., 2008{\natexlab{b}}, \apj, 674, 451

\bibitem[{{Burgasser} {et~al.}(2004){Burgasser}, {McElwain}, {Kirkpatrick},
  {Cruz}, {Tinney}, \& {Reid}}]{burgasser04a}
{Burgasser} A.~J., {McElwain} M.~W., {Kirkpatrick} J.~D., {Cruz} K.~L.,
  {Tinney} C.~G., {Reid} I.~N., 2004, \aj, 127, 2856

\bibitem[{{Burgasser} {et~al.}(2008{\natexlab{c}}){Burgasser}, {Tinney},
  {Cushing}, {Saumon}, {Marley}, {Bennett}, \& {Kirkpatrick}}]{burgasser08a}
{Burgasser} A.~J., {Tinney} C.~G., {Cushing} M.~C., {Saumon} D., {Marley}
  M.~S., {Bennett} C.~S., {Kirkpatrick} J.~D., 2008{\natexlab{c}}, \apjl, 689,
  L53

\bibitem[{{Burningham} {et~al.}(2008){Burningham}, {Pinfield}, {Leggett},
  {Tamura}, {Lucas}, {Homeier}, {Day-Jones}, {Jones}, {Clarke}, {Ishii},
  {Kuzuhara}, {Lodieu}, {Zapatero Osorio}, {Venemans}, {Mortlock}, {Barrado Y
  Navascu{\'e}s}, {Martin}, \& {Magazz{\`u}}}]{ben08}
{Burningham} B., {Pinfield} D.~J., {Leggett} S.~K., {Tamura} M., {Lucas} P.~W.,
  {Homeier} D., {Day-Jones} A., {Jones} H.~R.~A., {Clarke} J.~R.~A., {Ishii}
  M., {Kuzuhara} M., {Lodieu} N., {Zapatero Osorio} M.~R., {Venemans} B.~P.,
  {Mortlock} D.~J., {Barrado Y Navascu{\'e}s} D., {Martin} E.~L., {Magazz{\`u}}
  A., 2008, \mnras, 391, 320

\bibitem[{{Burningham} {et~al.}(2009){Burningham}, {Pinfield}, {Leggett},
  {Tinney}, {Liu}, {Homeier}, {West}, {Day-Jones}, {Huelamo}, {Dupuy}, {Zhang},
  {Murray}, {Lodieu}, {Barrado Y Navascu{\'e}s}, {Folkes}, {Galvez-Ortiz},
  {Jones}, {Lucas}, {Calderon}, \& {Tamura}}]{ben09}
{Burningham} B., {Pinfield} D.~J., {Leggett} S.~K., {Tinney} C.~G., {Liu}
  M.~C., {Homeier} D., {West} A.~A., {Day-Jones} A., {Huelamo} N., {Dupuy}
  T.~J., {Zhang} Z., {Murray} D.~N., {Lodieu} N., {Barrado Y Navascu{\'e}s} D.,
  {Folkes} S., {Galvez-Ortiz} M.~C., {Jones} H.~R.~A., {Lucas} P.~W.,
  {Calderon} M.~M., {Tamura} M., 2009, \mnras, 395, 1237

\bibitem[{{Burningham} {et~al.}(2010){Burningham}, {Pinfield}, {Lucas},
  {Leggett}, {Deacon}, {Tinney}, {Tamura}, {Lodieu}, {Zhang}, {Huelamo},
  {Murray}, {Mortlock}, {Barrado Y Navascu{\'e}s}, {Zapatero Osorio}, \&
  {Ishii}}]{ben10}
{Burningham} B., {Pinfield} D.~J., {Lucas} P.~W., {Leggett} S.~K., {Deacon}
  N.~R., {Tinney} C.~G., {Tamura} M., {Lodieu} N., {Zhang} Z., {Huelamo} N.,
  {Murray} D.~N., {Mortlock} D.~J., {Barrado Y Navascu{\'e}s} D., {Zapatero
  Osorio} M.~R., {Ishii} M., 2010, MNRAS submitted

\bibitem[{{Burrows} {et~al.}(2006){Burrows}, {Sudarsky}, \& {Hubeny}}]{bsh2006}
{Burrows} A., {Sudarsky} D., {Hubeny} I., 2006, \apj, 640, 1063

\bibitem[{{Casali} {et~al.}(2007){Casali}, {Adamson}, {Alves de Oliveira},
  {Almaini}, {Burch}, {Chuter}, {Elliot}, {Folger}, {Foucaud}, {Hambly},
  {Hastie}, {Henry}, {Hirst}, {Irwin}, {Ives}, {Lawrence}, {Laidlaw}, {Lee},
  {Lewis}, {Lunney}, {McLay}, {Montgomery}, {Pickup}, {Read}, {Rees}, {Robson},
  {Sekiguchi}, {Vick}, {Warren}, \& {Woodward}}]{wfcam}
{Casali} M., {Adamson} A., {Alves de Oliveira} C., {Almaini} O., {Burch} K.,
  {Chuter} T., {Elliot} J., {Folger} M., {Foucaud} S., {Hambly} N., {Hastie}
  M., {Henry} D., {Hirst} P., {Irwin} M., {Ives} D., {Lawrence} A., {Laidlaw}
  K., {Lee} D., {Lewis} J., {Lunney} D., {McLay} S., {Montgomery} D., {Pickup}
  A., {Read} M., {Rees} N., {Robson} I., {Sekiguchi} K., {Vick} A., {Warren}
  S., {Woodward} B., 2007, \aap, 467, 777

\bibitem[{{Cruz} {et~al.}(2007){Cruz}, {Reid}, {Kirkpatrick}, {Burgasser},
  {Liebert}, {Solomon}, {Schmidt}, {Allen}, {Hawley}, \& {Covey}}]{cruz07}
{Cruz} K.~L., {Reid} I.~N., {Kirkpatrick} J.~D., {Burgasser} A.~J., {Liebert}
  J., {Solomon} A.~R., {Schmidt} S.~J., {Allen} P.~R., {Hawley} S.~L., {Covey}
  K.~R., 2007, \aj, 133, 439

\bibitem[{{Cushing} {et~al.}(2009){Cushing}, {Looper}, {Burgasser},
  {Kirkpatrick}, {Faherty}, {Cruz}, {Sweet}, \& {Sanderson}}]{cushing09}
{Cushing} M.~C., {Looper} D., {Burgasser} A.~J., {Kirkpatrick} J.~D., {Faherty}
  J., {Cruz} K.~L., {Sweet} A., {Sanderson} R.~E., 2009, \apj, 696, 986

\bibitem[{{Cushing} {et~al.}(2008){Cushing}, {Marley}, {Saumon}, {Kelly},
  {Vacca}, {Rayner}, {Freedman}, {Lodders}, \& {Roellig}}]{cushing08}
{Cushing} M.~C., {Marley} M.~S., {Saumon} D., {Kelly} B.~C., {Vacca} W.~D.,
  {Rayner} J.~T., {Freedman} R.~S., {Lodders} K., {Roellig} T.~L., 2008, \apj,
  678, 1372

\bibitem[{{Delfosse} {et~al.}(1997){Delfosse}, {Tinney}, {Forveille},
  {Epchtein}, {Bertin}, {Borsenberger}, {Copet}, {de Batz}, {Fouque},
  {Kimeswenger}, {Le Bertre}, {Lacombe}, {Rouan}, \& {Tiphene}}]{delfosse97}
{Delfosse} X., {Tinney} C.~G., {Forveille} T., {Epchtein} N., {Bertin} E.,
  {Borsenberger} J., {Copet} E., {de Batz} B., {Fouque} P., {Kimeswenger} S.,
  {Le Bertre} T., {Lacombe} F., {Rouan} D., {Tiphene} D., 1997, \aap, 327, L25

\bibitem[{{Delorme} {et~al.}(2008{\natexlab{a}}){Delorme}, {Delfosse},
  {Albert}, {Artigau}, {Forveille}, {Reyl{\'e}}, {Allard}, {Homeier}, {Robin},
  {Willott}, {Liu}, \& {Dupuy}}]{delorme08}
{Delorme} P., {Delfosse} X., {Albert} L., {Artigau} E., {Forveille} T.,
  {Reyl{\'e}} C., {Allard} F., {Homeier} D., {Robin} A.~C., {Willott} C.~J.,
  {Liu} M.~C., {Dupuy} T.~J., 2008{\natexlab{a}}, \aap, 482, 961

\bibitem[{{Delorme} {et~al.}(2008{\natexlab{b}}){Delorme}, {Willott},
  {Forveille}, {Delfosse}, {Reyl{\'e}}, {Bertin}, {Albert}, {Artigau}, {Robin},
  {Allard}, {Doyon}, \& {Hill}}]{cfbds}
{Delorme} P., {Willott} C.~J., {Forveille} T., {Delfosse} X., {Reyl{\'e}} C.,
  {Bertin} E., {Albert} L., {Artigau} E., {Robin} A.~C., {Allard} F., {Doyon}
  R., {Hill} G.~J., 2008{\natexlab{b}}, \aap, 484, 469

\bibitem[{{Geballe} {et~al.}(2002){Geballe}, {Knapp}, {Leggett}, {Fan},
  {Golimowski}, {Anderson}, {Brinkmann}, {Csabai}, {Gunn}, {Hawley},
  {Hennessy}, {Henry}, {Hill}, {Hindsley}, {Ivezi{\'c}}, {Lupton}, {McDaniel},
  {Munn}, {Narayanan}, {Peng}, {Pier}, {Rockosi}, {Schneider}, {Smith},
  {Strauss}, {Tsvetanov}, {Uomoto}, {York}, \& {Zheng}}]{geballe02}
{Geballe} T.~R., {Knapp} G.~R., {Leggett} S.~K., {Fan} X., {Golimowski} D.~A.,
  {Anderson} S., {Brinkmann} J., {Csabai} I., {Gunn} J.~E., {Hawley} S.~L.,
  {Hennessy} G., {Henry} T.~J., {Hill} G.~J., {Hindsley} R.~B., {Ivezi{\'c}}
  {\v Z}., {Lupton} R.~H., {McDaniel} A., {Munn} J.~A., {Narayanan} V.~K.,
  {Peng} E., {Pier} J.~R., {Rockosi} C.~M., {Schneider} D.~P., {Smith} J.~A.,
  {Strauss} M.~A., {Tsvetanov} Z.~I., {Uomoto} A., {York} D.~G., {Zheng} W.,
  2002, \apj, 564, 466

\bibitem[{{Geballe} {et~al.}(2009){Geballe}, {Saumon}, {Golimowski}, {Leggett},
  {Marley}, \& {Noll}}]{geballe09}
{Geballe} T.~R., {Saumon} D., {Golimowski} D.~A., {Leggett} S.~K., {Marley}
  M.~S., {Noll} K.~S., 2009, \apj, 695, 844

\bibitem[{{Gizis}(1997)}]{gizis97}
{Gizis} J.~E., 1997, \aj, 113, 806

\bibitem[{{Hambly} {et~al.}(2008){Hambly}, {Collins}, {Cross}, {Mann}, {Read},
  {Sutorius}, {Bond}, {Bryant}, {Emerson}, {Lawrence}, {Rimoldini}, {Stewart},
  {Williams}, {Adamson}, {Hirst}, {Dye}, \& {Warren}}]{wsa}
{Hambly} N.~C., {Collins} R.~S., {Cross} N.~J.~G., {Mann} R.~G., {Read} M.~A.,
  {Sutorius} E.~T.~W., {Bond} I., {Bryant} J., {Emerson} J.~P., {Lawrence} A.,
  {Rimoldini} L., {Stewart} J.~M., {Williams} P.~M., {Adamson} A., {Hirst} P.,
  {Dye} S., {Warren} S.~J., 2008, \mnras, 384, 637

\bibitem[{{Hambly} {et~al.}(2001){Hambly}, {MacGillivray}, {Read}, {Tritton},
  {Thomson}, {Kelly}, {Morgan}, {Smith}, {Driver}, {Williamson}, {Parker},
  {Hawkins}, {Williams}, \& {Lawrence}}]{supercos1}
{Hambly} N.~C., {MacGillivray} H.~T., {Read} M.~A., {Tritton} S.~B., {Thomson}
  E.~B., {Kelly} B.~D., {Morgan} D.~H., {Smith} R.~E., {Driver} S.~P.,
  {Williamson} J., {Parker} Q.~A., {Hawkins} M.~R.~S., {Williams} P.~M.,
  {Lawrence} A., 2001, \mnras, 326, 1279

\bibitem[{{Haywood} {et~al.}(1997){Haywood}, {Robin}, \& {Creze}}]{haywood97}
{Haywood} M., {Robin} A.~C., {Creze} M., 1997, \aap, 320, 428

\bibitem[{{Irwin} {et~al.}(2004){Irwin}, {Lewis}, {Hodgkin}, {Bunclark},
  {Evans}, {McMahon}, {Emerson}, {Stewart}, \& {Beard}}]{irwin04}
{Irwin} M.~J., {Lewis} J., {Hodgkin} S., {Bunclark} P., {Evans} D., {McMahon}
  R., {Emerson} J.~P., {Stewart} M., {Beard} S., 2004, in Society of
  Photo-Optical Instrumentation Engineers (SPIE) Conference Series, Vol. 5493,
  Society of Photo-Optical Instrumentation Engineers (SPIE) Conference Series,
  {P.~J.~Quinn \& A.~Bridger}, ed., pp. 411--422

\bibitem[{{Kirkpatrick}(2005)}]{kirkpatrick05}
{Kirkpatrick} J.~D., 2005, \araa, 43, 195

\bibitem[{{Kirkpatrick} {et~al.}(1999){Kirkpatrick}, {Reid}, {Liebert},
  {Cutri}, {Nelson}, {Beichman}, {Dahn}, {Monet}, {Gizis}, \&
  {Skrutskie}}]{kirkpatrick99}
{Kirkpatrick} J.~D., {Reid} I.~N., {Liebert} J., {Cutri} R.~M., {Nelson} B.,
  {Beichman} C.~A., {Dahn} C.~C., {Monet} D.~G., {Gizis} J.~E., {Skrutskie}
  M.~F., 1999, \apj, 519, 802

\bibitem[{{Kirkpatrick} {et~al.}(2000){Kirkpatrick}, {Reid}, {Liebert},
  {Gizis}, {Burgasser}, {Monet}, {Dahn}, {Nelson}, \&
  {Williams}}]{kirkpatrick2000}
{Kirkpatrick} J.~D., {Reid} I.~N., {Liebert} J., {Gizis} J.~E., {Burgasser}
  A.~J., {Monet} D.~G., {Dahn} C.~C., {Nelson} B., {Williams} R.~J., 2000, \aj,
  120, 447

\bibitem[{{Knapp} {et~al.}(2004){Knapp}, {Leggett}, {Fan}, {Marley}, {Geballe},
  {Golimowski}, {Finkbeiner}, {Gunn}, {Hennawi}, {Ivezi{\'c}}, {Lupton},
  {Schlegel}, {Strauss}, {Tsvetanov}, {Chiu}, {Hoversten}, {Glazebrook},
  {Zheng}, {Hendrickson}, {Williams}, {Uomoto}, {Vrba}, {Henden}, {Luginbuhl},
  {Guetter}, {Munn}, {Canzian}, {Schneider}, \& {Brinkmann}}]{knapp04}
{Knapp} G.~R., {Leggett} S.~K., {Fan} X., {Marley} M.~S., {Geballe} T.~R.,
  {Golimowski} D.~A., {Finkbeiner} D., {Gunn} J.~E., {Hennawi} J., {Ivezi{\'c}}
  Z., {Lupton} R.~H., {Schlegel} D.~J., {Strauss} M.~A., {Tsvetanov} Z.~I.,
  {Chiu} K., {Hoversten} E.~A., {Glazebrook} K., {Zheng} W., {Hendrickson} M.,
  {Williams} C.~C., {Uomoto} A., {Vrba} F.~J., {Henden} A.~A., {Luginbuhl}
  C.~B., {Guetter} H.~H., {Munn} J.~A., {Canzian} B., {Schneider} D.~P.,
  {Brinkmann} J., 2004, \aj, 127, 3553

\bibitem[{{Kobayashi} {et~al.}(2000){Kobayashi}, {Tokunaga}, {Terada}, {Goto},
  {Weber}, {Potter}, {Onaka}, {Ching}, {Young}, {Fletcher}, {Neil},
  {Robertson}, {Cook}, {Imanishi}, \& {Warren}}]{IRCS2000}
{Kobayashi} N., {Tokunaga} A.~T., {Terada} H., {Goto} M., {Weber} M., {Potter}
  R., {Onaka} P.~M., {Ching} G.~K., {Young} T.~T., {Fletcher} K., {Neil} D.,
  {Robertson} L., {Cook} D., {Imanishi} M., {Warren} D.~W., 2000, in Presented
  at the Society of Photo-Optical Instrumentation Engineers (SPIE) Conference,
  Vol. 4008, Proc. SPIE Vol. 4008, p. 1056-1066, Optical and IR Telescope
  Instrumentation and Detectors, Masanori Iye; Alan F. Moorwood; Eds., {Iye}
  M., {Moorwood} A.~F., eds., pp. 1056--1066

\bibitem[{{Lafreni{\`e}re} {et~al.}(2008){Lafreni{\`e}re}, {Jayawardhana},
  {Brandeker}, {Ahmic}, \& {van Kerkwijk}}]{lafreniere08}
{Lafreni{\`e}re} D., {Jayawardhana} R., {Brandeker} A., {Ahmic} M., {van
  Kerkwijk} M.~H., 2008, \apj, 683, 844

\bibitem[{{Lawrence} {et~al.}(2007){Lawrence}, {Warren}, {Almaini}, {Edge},
  {Hambly}, {Jameson}, {Lucas}, {Casali}, {Adamson}, \& {and thirteen
  co-authors}}]{ukidss}
{Lawrence} A., {Warren} S.~J., {Almaini} O., {Edge} A.~C., {Hambly} N.~C.,
  {Jameson} R.~F., {Lucas} P., {Casali} M., {Adamson} A., {and thirteen
  co-authors}, 2007, \mnras, 379, 1599

\bibitem[{{Leggett} {et~al.}(2010){Leggett}, {Burningham}, {Saumon}, {Marley},
  {Warren}, {Smart}, {Jones}, {Lucas}, {Pinfield}, \& {Tamura}}]{sandy10}
{Leggett} S.~K., {Burningham} B., {Saumon} D., {Marley} M.~S., {Warren} S.~J.,
  {Smart} R.~L., {Jones} H.~R.~A., {Lucas} P.~W., {Pinfield} D.~J., {Tamura}
  M., 2010, ArXiv e-prints

\bibitem[{{Leggett} {et~al.}(2009){Leggett}, {Cushing}, {Saumon}, {Marley},
  {Roellig}, {Warren}, {Burningham}, {Jones}, {Kirkpatrick}, {Lodieu}, {Lucas},
  {Mainzer}, {Mart{\'{\i}}n}, {McCaughrean}, {Pinfield}, {Sloan}, {Smart},
  {Tamura}, \& {Van Cleve}}]{sandy09}
{Leggett} S.~K., {Cushing} M.~C., {Saumon} D., {Marley} M.~S., {Roellig} T.~L.,
  {Warren} S.~J., {Burningham} B., {Jones} H.~R.~A., {Kirkpatrick} J.~D.,
  {Lodieu} N., {Lucas} P.~W., {Mainzer} A.~K., {Mart{\'{\i}}n} E.~L.,
  {McCaughrean} M.~J., {Pinfield} D.~J., {Sloan} G.~C., {Smart} R.~L., {Tamura}
  M., {Van Cleve} J., 2009, \apj, 695, 1517

\bibitem[{{Leggett} {et~al.}(2007){Leggett}, {Marley}, {Freedman}, {Saumon},
  {Liu}, {Geballe}, {Golimowski}, \& {Stephens}}]{sandy07}
{Leggett} S.~K., {Marley} M.~S., {Freedman} R., {Saumon} D., {Liu} M.~C.,
  {Geballe} T.~R., {Golimowski} D.~A., {Stephens} D.~C., 2007, \apj, 667, 537

\bibitem[{{Liebert} \& {Burgasser}(2007)}]{lb07}
{Liebert} J., {Burgasser} A.~J., 2007, \apj, 655, 522

\bibitem[{{Liu} {et~al.}(2007){Liu}, {Leggett}, \& {Chiu}}]{liu07}
{Liu} M.~C., {Leggett} S.~K., {Chiu} K., 2007, \apj, 660, 1507

\bibitem[{{Liu} {et~al.}(2006){Liu}, {Leggett}, {Golimowski}, {Chiu}, {Fan},
  {Geballe}, {Schneider}, \& {Brinkmann}}]{liu06}
{Liu} M.~C., {Leggett} S.~K., {Golimowski} D.~A., {Chiu} K., {Fan} X.,
  {Geballe} T.~R., {Schneider} D.~P., {Brinkmann} J., 2006, \apj, 647, 1393

\bibitem[{{Lodieu} {et~al.}(2007){Lodieu}, {Pinfield}, {Leggett}, {Jameson},
  {Mortlock}, {Warren}, {Burningham}, {Lucas}, {Chiu}, {Liu}, {Venemans},
  {McMahon}, {Allard}, {Baraffe}, {Y Navascu{\'e}s}, {Carraro}, {Casewell},
  {Chabrier}, {Chappelle}, {Clarke}, {Day-Jones}, {Deacon}, {Dobbie}, {Folkes},
  {Hambly}, {Hewett}, {Hodgkin}, {Jones}, {Kendall}, {Magazz{\`u}},
  {Mart{\'{\i}}n}, {McCaughrean}, {Nakajima}, {Pavlenko}, {Tamura}, {Tinney},
  \& {Zapatero Osorio}}]{lod07}
{Lodieu} N., {Pinfield} D.~J., {Leggett} S.~K., {Jameson} R.~F., {Mortlock}
  D.~J., {Warren} S.~J., {Burningham} B., {Lucas} P.~W., {Chiu} K., {Liu}
  M.~C., {Venemans} B.~P., {McMahon} R.~G., {Allard} F., {Baraffe} I., {Y
  Navascu{\'e}s} D.~B., {Carraro} G., {Casewell} S.~L., {Chabrier} G.,
  {Chappelle} R.~J., {Clarke} F., {Day-Jones} A.~C., {Deacon} N.~R., {Dobbie}
  P.~D., {Folkes} S.~L., {Hambly} N.~C., {Hewett} P.~C., {Hodgkin} S.~T.,
  {Jones} H.~R.~A., {Kendall} T.~R., {Magazz{\`u}} A., {Mart{\'{\i}}n} E.~L.,
  {McCaughrean} M.~J., {Nakajima} T., {Pavlenko} Y., {Tamura} M., {Tinney}
  C.~G., {Zapatero Osorio} M.~R., 2007, \mnras, 379, 1423

\bibitem[{{Lodieu} {et~al.}(2009){Lodieu}, {Zapatero Osorio}, {Martin},
  {Solano}, \& {Aberaturi}}]{lodieu2009}
{Lodieu} N., {Zapatero Osorio} M.~R., {Martin} E.~L., {Solano} E., {Aberaturi}
  M., 2009, ArXiv e-prints

\bibitem[{{Marley} {et~al.}(2002){Marley}, {Seager}, {Saumon}, {Lodders},
  {Ackerman}, {Freedman}, \& {Fan}}]{marley02}
{Marley} M.~S., {Seager} S., {Saumon} D., {Lodders} K., {Ackerman} A.~S.,
  {Freedman} R.~S., {Fan} X., 2002, \apj, 568, 335

\bibitem[{{Pinfield} {et~al.}(2008){Pinfield}, {Burningham}, {Tamura},
  {Leggett}, {Lodieu}, {Lucas}, {Mortlock}, {Warren}, {Homeier}, {Ishii},
  {McMahon}, \& {and 29 co-authors}}]{pinfield08}
{Pinfield} D.~J., {Burningham} B., {Tamura} M., {Leggett} S.~K., {Lodieu} N.,
  {Lucas} P.~W., {Mortlock} D.~J., {Warren} S.~J., {Homeier} D., {Ishii} M.,
  {McMahon} R.~G., {and 29 co-authors}, 2008, MNRAS submitted

\bibitem[{{Reach} {et~al.}(2005){Reach}, {Megeath}, {Cohen}, {Hora}, {Carey},
  {Surace}, {Willner}, {Barmby}, {Wilson}, {Glaccum}, {Lowrance}, {Marengo}, \&
  {Fazio}}]{reach05}
{Reach} W.~T., {Megeath} S.~T., {Cohen} M., {Hora} J., {Carey} S., {Surace} J.,
  {Willner} S.~P., {Barmby} P., {Wilson} G., {Glaccum} W., {Lowrance} P.,
  {Marengo} M., {Fazio} G.~G., 2005, \pasp, 117, 978

\bibitem[{{Reiners} \& {Basri}(2006)}]{reiners06}
{Reiners} A., {Basri} G., 2006, \aj, 131, 1806

\bibitem[{{Robin} {et~al.}(2003){Robin}, {Reyl{\'e}}, {Derri{\`e}re}, \&
  {Picaud}}]{robin03}
{Robin} A.~C., {Reyl{\'e}} C., {Derri{\`e}re} S., {Picaud} S., 2003, \aap, 409,
  523

\bibitem[{{Saumon} {et~al.}(1994){Saumon}, {Bergeron}, {Lunine}, {Hubbard}, \&
  {Burrows}}]{saumon94}
{Saumon} D., {Bergeron} P., {Lunine} J.~I., {Hubbard} W.~B., {Burrows} A.,
  1994, \apj, 424, 333

\bibitem[{{Saumon} \& {Marley}(2008)}]{sm08}
{Saumon} D., {Marley} M.~S., 2008, \apj, 689, 1327

\bibitem[{{Schilbach} {et~al.}(2009){Schilbach}, {R{\"o}ser}, \&
  {Scholz}}]{schilbach09}
{Schilbach} E., {R{\"o}ser} S., {Scholz} R., 2009, \aap, 493, L27

\bibitem[{{Schmidt} {et~al.}(2009){Schmidt}, {West}, {Burgasser}, {Bochanski},
  \& {Hawley}}]{schmidt10}
{Schmidt} S.~J., {West} A.~A., {Burgasser} A.~J., {Bochanski} J.~J., {Hawley}
  S.~L., 2009, ApJ in press, ArXiv 0912.3565

\bibitem[{{Sivarani} {et~al.}(2009){Sivarani}, {L{\'e}pine}, {Kembhavi}, \&
  {Gupchup}}]{sivarani09}
{Sivarani} T., {L{\'e}pine} S., {Kembhavi} A.~K., {Gupchup} J., 2009, \apjl,
  694, L140

\bibitem[{{Stephens} \& {Leggett}(2004)}]{stephens04}
{Stephens} D.~C., {Leggett} S.~K., 2004, \pasp, 116, 9

\bibitem[{{Stephens} {et~al.}(2009){Stephens}, {Leggett}, {Cushing}, {Marley},
  {Saumon}, {Geballe}, {Golimowski}, {Fan}, \& {Noll}}]{stephens09}
{Stephens} D.~C., {Leggett} S.~K., {Cushing} M.~C., {Marley} M.~S., {Saumon}
  D., {Geballe} T.~R., {Golimowski} D.~A., {Fan} X., {Noll} K.~S., 2009, \apj,
  702, 154

\bibitem[{{Tokunaga} {et~al.}(2002){Tokunaga}, {Simons}, \& {Vacca}}]{mko}
{Tokunaga} A.~T., {Simons} D.~A., {Vacca} W.~D., 2002, \pasp, 114, 180

\bibitem[{{Warren} {et~al.}(2007){Warren}, {Mortlock}, {Leggett}, {Pinfield},
  {Homeier}, {Dye}, {Jameson}, {Lodieu}, {Lucas}, {Adamson}, \& {and 14
  co-authors}}]{warren07}
{Warren} S.~J., {Mortlock} D.~J., {Leggett} S.~K., {Pinfield} D.~J., {Homeier}
  D., {Dye} S., {Jameson} R.~F., {Lodieu} N., {Lucas} P.~W., {Adamson} A.~J.,
  {and 14 co-authors}, 2007, \mnras, 381, 1400

\end{thebibliography}

\end{document}